\documentclass[12pt]{iopart}
\expandafter\let\csname equation*\endcsname\relax
\expandafter\let\csname endequation*\endcsname\relax 
\usepackage{mathtools} 

\usepackage{natbib}
\usepackage{cite}
\usepackage{algorithmic}
\usepackage{graphicx}
\usepackage{textcomp}
\usepackage{xcolor}
\usepackage[font=small,labelfont=bf]{caption}
\usepackage{float}

\usepackage{fancyhdr}
\usepackage[margin=2.5cm,top=2.5cm,headheight=16pt,headsep=0.50in,heightrounded]{geometry}

\usepackage{lineno}

\fancypagestyle{FirstPage}{
\lhead{Preprint version. Manuscript submitted to the Journal of Physics in Medicine and Biology}
\rhead{}
\setlength{\headheight}{20pt}
}

\def\argmin#1{\protect\mathop{\mathrm{arg\,min}}_{#1}}

\begin{document}

\title{CNN-Based PET Sinogram Repair to Mitigate Defective Block Detectors}
\thispagestyle{FirstPage}

\author{William Whiteley$^{1,2}$, Jens Gregor$^1$}

\address{$^1$ The University of Tennessee, Knoxville, TN 37996}
\address{$^2$ Siemens Medical Solutions USA Inc., Knoxville, TN, 37932}
\ead{wwhitele@vols.utk.edu}
\vspace{10pt}
\begin{indented}
\item[]July 4, 2019
\end{indented}

\begin{abstract}
Positron emission tomography (PET) scanners continue to increase sensitivity and axial coverage by adding an ever expanding array of block detectors. As they age, one or more block detectors may lose sensitivity due to a malfunction or component failure. The sinogram data missing as a result thereof can lead to 
artifacts and other image degradations. We propose to mitigate the effects of malfunctioning block detectors
by carrying out sinogram repair using a deep convolutional neural network. Experiments using whole-body patient studies with varying amounts of raw data removed are used to show
that the neural network significantly outperforms previously published methods with respect to normalized mean squared error for raw sinograms, a multi-scale structural similarity measure for reconstructed images and with regard to quantitative accuracy.
\end{abstract}

\section{Introduction}
\begin{figure}[t]
	\centering
	\includegraphics[width=0.7\columnwidth]{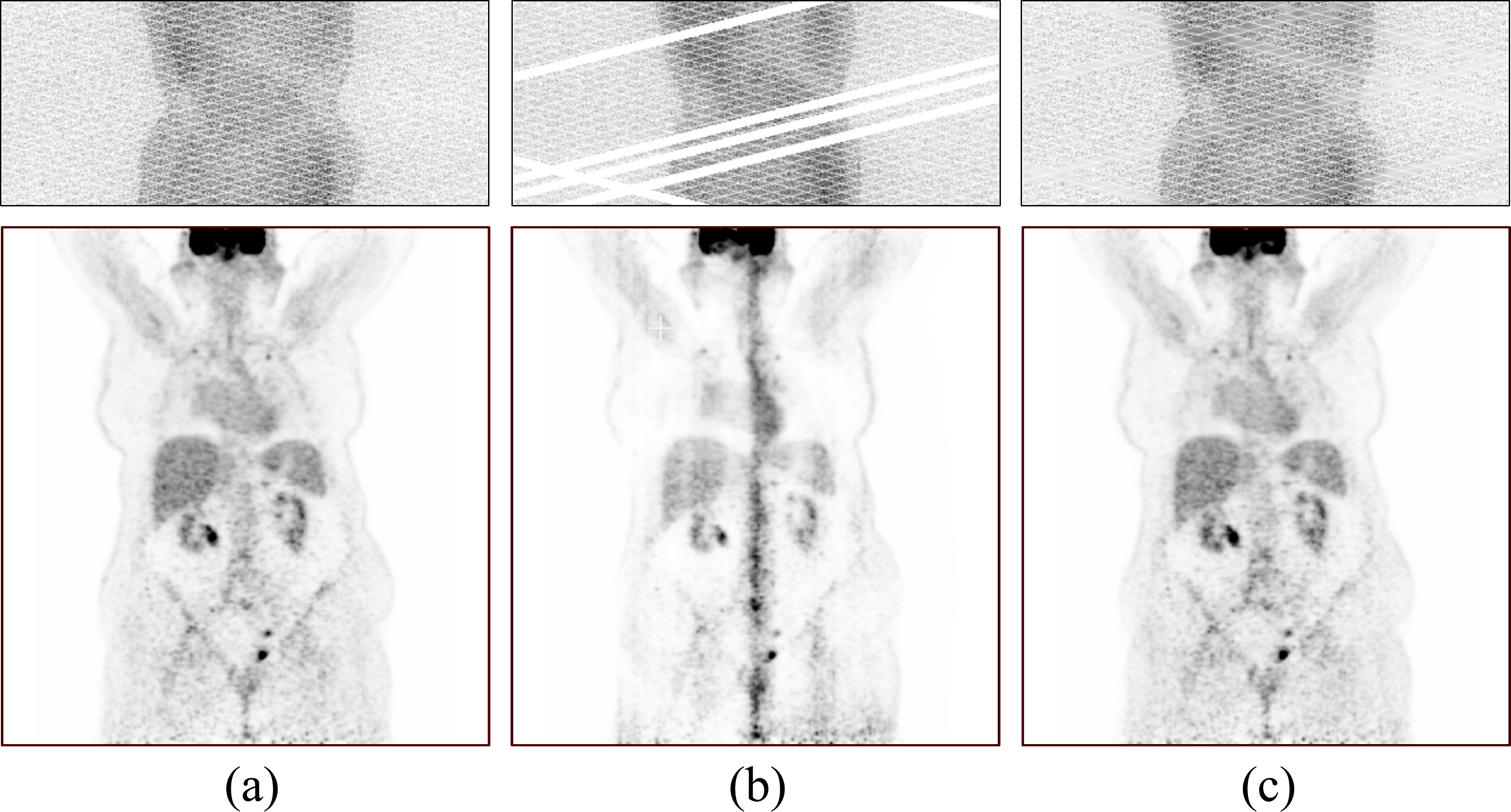}
	\caption{
	  Illustration of the problem and the proposed solution:
	  (top row) PET sinograms and (bottom row) reconstructed images;
	  (a) standard clinical quality data; 
	  (b) effect of malfunctioning detector blocks; 
	  and
	  (c) deep neural network restored data.
	}
	\label{initial}
\end{figure}

The combination of positron emission tomography and X-ray computed tomography (PET/CT) 
is used extensively both in both clinical and research environments where device up-time is a key operational metric for imaging organizations. During a PET scanner's life, however, one or more block detectors in the imaging array will likely malfunction causing the system to lose coincidence events along the corresponding lines of response (LORs). As illustrated in Fig.~\ref{initial}, this creates gaps in the sinogram data which leads to artifacts and image degradation, as well as downtime for the scanner \citep{Zito:07,Elhami:2010,Voert:16}. This scenario is likely to become even more common as technological advances lead to scanners with longer axial fields-of-view such as the development and commercialization of whole-body PET scanners \citep{Cherry2018}. This paper presents a deep learning based technique to mitigate single and multiple failed block detectors, while restoring reconstructed image quality.

Previous research on mitigating missing PET data from either intentional gaps due to the design of the PET scanner or unintentional regions of missing data due to the failure one or more block detectors largely focused on interpolation, model based or transform based methods. Sinogram repair using interpolation with linear, bilinear and bicubic methods along the radial, angular or axial directions is computationally efficient and reduces reconstructed image artifacts considerably \citep{Jong:03}, but has been shown to have lower image quality than model \citep{Baghaei:08} and transform \citep{Velden:08} based methods. A more advanced interpolation method comparable to the model based and transform methods involves collapsing the transradial slices where data is missing followed by bicubic interpolation based resizing \citep{Peltonen:13}. The model based methods use either a statistical \citep{Kinahan:97,Baghaei:08,Velden:08} or analytical \citep{Baghaei:00} reconstruction algorithm to create an initial image estimate from a sinogram with missing data and then forward project these images using the PET system model to estimate the missing data repeating the process iteratively. Transform based repair methods convert a sinogram from the spatial domain into the frequency domain and then filter the frequency domain coefficients corresponding to the missing data. The filtered data is then transformed back into the spatial domain and the process is repeated until a stopping criterion is satisfied. The constrained Fourier space (CFS) method \citep{karp:88}, which employs a bow-tie shaped frequency filter, was the first transform based PET sinogram repair method widely studied \citep{Velden:08, Baghaei:08,Lehnert:06, Buchert:99}. The discrete cosine transform (DCT) was also explored for this purpose \citep{Tuna:08} demonstrating promising results using a wedge shaped frequency filter.

There is little published research on deep learning based PET sinogram repair methods, but techniques similar to the methods proposed in this study are used in medical imaging for limited angle / low dose x-ray CT \citep{Kan:16,Anirudh:18,Xie:2018,zhao:2018,ziheng:19} and artifact mitigation \citep{zhang:18}. Inpainting is a general topic of interest in the wider deep learning literature beginning with the work of Xie, Xu and Chen \citep{Xie:2012} who applied a stacked sparse auto-encoder to the problem and this was followed by numerous additional works that primarily applied generative adversarial networks (GANs) \citep{Goodfellow:14} to the inpainting problem such as \citep{Yeh:17,yang:17, Wang:18, Yu:2018}. While we considered using a GAN for our proposed network, we decided the additional complexity was not necessary and opted for a simpler approach. 

In a deep learning context, sinogram repair can be framed as mapping input sinograms $x$ with missing LORs to repaired sinograms $\hat{y}$ by learning an inpainting neural network function $f(x;\theta)=\hat{y}$ parameterized by weights $\theta$ and trained by minimizing a loss function $\mathcal{L}$ over a training data set $(x_n, y_n)$ with back-propagation and gradient descent \citep{yann:88}.
The solution is given by

\begin{align}
      \theta^*= \argmin{\theta} \frac{1}{n}\sum_{i=1}^n \mathcal{L}(f(x_i;\theta),y_i).
\end{align}
The performance of the Sinogram Repair Network presented in this paper for implementing $f(x;\theta^*)$
was evaluated with whole-body patient data for one, two and four missing block detectors.
Results were quantitatively compared in sinogram space and image space to the previously published sinogram repair methods of interpolation, the constrained Fourier space method and a model based method utilizing iterative reconstruction.

\section{Materials and methods}

\subsection{Sinogram Repair Network}
\begin{figure*}[ht]
	\centering
	\includegraphics[width=6.0in]{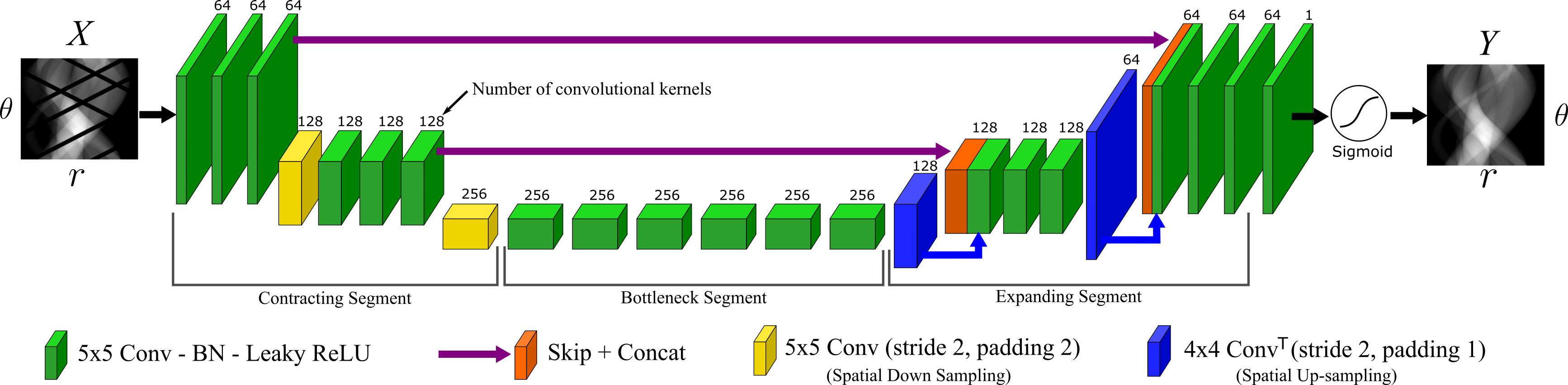}
	\caption{Each green block represents a layer containing convolution, batch normalization (BN) and leaky-ReLU activation. The purple arrows and orange blocks represent skip connections and concatenation of the features from the contracting layers with the expansive layers. The network up-samples using transpose convolution to double the spatial dimensions. The sigmoid activation function at the end scales the output and acts as a stabilizing element. The number above each block indicates the number of kernels in that layer. 
	  }
	\label{arch}
\end{figure*}

Illustrated in Fig.~\ref{arch}, our proposed neural network is a variant of the well known U-NET architecture \citep{Ronneberger:15}. The network consists of 23 convolutional layers with a contracting segment (8 layers) that extracts positional features and twice performs spatial down-sampling, a bottle neck segment (6 layers) that learns a compressed representation, and an expansive segment (8 layers) that spatially up-samples the compressed representation and extracts contextual information. The contracting and expansive segments are connected by skip connections that allow the network to combine the positional and contextual information to predict the missing sinogram information. The final sigmoid activation function helps provide stability during training.  

All convolutional kernels are 5x5 except for the up-sampling layers that use 4x4 kernels. Padding is employed to ensure that the dimensions remain consistent after each convolution. Down-sampling to half-size in each dimension is accomplished by using a kernel stride of 2, and up-sampling is accomplished by using a fractionally strided convolution (also known as transposed convolution). After each spatial down-sampling, the number of kernels in a convolutional layer is doubled, and with each up-sampling the number of kernels in a convolutional layer is halved. Each convolutional layer is followed by batch normalization and a Leaky-ReLU activation function with a negative slope of 0.2.

\subsection{Experimental Data Sets}

The training and test data was acquired using a Siemens Biograph mCT scanner from 20 PET/CT patient scans using a whole-body non-time-of-flight protocol of normal clinical duration with approximately 2-3 minutes per bed position. Due to the varying distribution of the radioactive source and attenuating structures of the body, these types of PET scans produce a wide distribution of sinograms from sparse and noisy containing less than 20,000 counts up to densely packed sinograms with rich structures containing well over 300,000 counts. The resulting data set contained 66,798 raw (uncorrected) sinogram slices counting all direct and oblique planes. From this data set 2 randomly selected patients totalling 7,164 sinograms became the test set with the remaining slices used to train the Sinogram Repair Network. To augment the data set even further during training, the input slices were randomly flipped horizontally and vertically with a probability of 0.5. Since the original data set does not contain missing information, defective block detectors were simulated by generating a set of sinogram masks, one for each of the 48 block detectors in the PET gantry and randomly applying them during training. Although based on the original 59,634 training sinograms, this dynamic augmentation and random removal of LORs meant the neural network rarely, if ever, saw the same data sample during training, which significantly aided in generalization and preventing overfitting. With the 7,164 slices in the test set, 10 realizations of missing blocks were created for each of one, two, and four missing blocks with no additional augmentation.  

Since the range of potential values in a sinogram are only limited by the size of the integer representing a given bin, which in this case is 16 bit, instability can occur during neural network training. To control this unbounded parameter each sinogram is scaled down to a range $[0:1]$ prior to input to the network and the output is scaled back up using the same scaling factor. Scaling is an important aspect to consider when working with sinogram data sets since measurements such as Standardized Uptake Values (SUV) are important quantitative diagnostic measures taken from the resulting images.

\subsection{Network Training}

The proposed Sinogram Repair Network was implemented using PyTorch \citep{paszke:2017} and trained for 4,500 epochs with each epoch utilizing 2,048 sinogram slices randomly sampled from the training set without replacement in mini-batches of 32. At runtime, randomly selected masks were applied to the sampled sinograms to create missing data. The Adam optimizer \citep{Kingma:14} was employed, which is an extension to classic stochastic gradient descent that maintains a learning rate for each network parameter separately, with an initial learning rate of 0.0001, a first momentum of 0.5 and a second momentum of 0.999. After each $n$ epochs, a warm restart \citep{Loshchilov:16} was performed where the learning rate was reset to its initial value and the optimizer's momentum buffers were cleared creating a temporary period of instability allowing the optimizer to escape sub-optimal local minimums. The value for $n$ was incremented by a value $\Delta n$ at each reset to increase the duration between warm restarts. We set the initial value of $n$ to 300 and $\Delta n$ to 50. Various loss functions were explored and we chose to follow previous research dedicated to loss functions for image repair \citep{Zhao:17} and use a weighted combination of an element-wise L1 and multi-scale structural similarity (MS-SSIM) \citep{Wang:03} based loss. Specifically,

\begin{align}
\mathcal{L}(\hat{y},y) &= 
\alpha \ \frac{1}{k}\sum_{i=0}^{k-1}|\hat{y}_i-y_i| +
\beta \ l_M(\hat{y},y) \prod_{j=1}^M c_j(\hat{y},y) s_j(\hat{y},y)
\end{align}
where each sinogram $\hat{y}$ and $y$ contains $k$ bins and the SSIM components are calculated on a sliding 11 x 11 window between sinograms $\hat{y}$ and $y$ where $\mu$ is the mean, $\sigma$ is the variance and $\sigma_{\hat{y}y}$ is the covariance of $\hat{y}$ and $y$.
\begin{align}
l_M(\hat{y},y) = \frac{2\mu_{\hat{y}}\mu_y+C_1}{\mu_{\hat{y}}^2+\mu_{y}^2+C_1}    
\end{align}
\begin{align}
c_j(\hat{y},y) = \frac{2\sigma_{\hat{y}}\sigma_y+C_2}{\sigma_{\hat{y}}^2+\sigma_{y}^2+C_2}    
\end{align}
\begin{align}
s_j(\hat{y},y) = \frac{\sigma_{\hat{y}y}+C_3}{\sigma_{\hat{y}}\sigma_y+C_3}    
\end{align}
where $C_1$,$C_2$ and $C_3$ are constants given by
\begin{align}
    C_1=(K_1 L)^2 \ , \ C_2=(K_2 L)^2 \ , \ C_3=C_2/2 \ ,
\end{align}
and $L$ is the dynamic range of values (1 in our case), $K_1=0.01$ and $K_2=0.03$. The luminance component ($l_M$) is only evaluated at the highest scale $M$ whereas the contrast ($c_j$) and structure ($s_j$) measures are evaluated at five scales each half the dimension of the previous. The values of $\alpha$ and $\beta$ were explored empirically with the results reported below based on $\alpha \!=\! 1$ and $\beta \!=\! 0.75$.

\subsection{Evaluation Methods}

The Sinogram Repair Network (SRN) performance is evaluated in sinogram, image, and quantitative accuracy domains. Comparisons are made against each of the top performing previously established repair methods, namely, interpolation, model based and transform based. For interpolation, the method described in \citep{Peltonen:13} is used. The model based method uses OSEM reconstruction from the TomoPy library \citep{Gursoy:14} with 10 iterations, and the transform based method uses the constrained Fourier spaced (CFS) method \citep{karp:88} with a bow-tie filter and 100 iterations.        

In the sinogram domain, ground truth samples were compared to repaired sinograms for each of the repair methods across all 7,164 slices in the test data set with one, two and four missing blocks. Since all repair methods fill in the missing information in the original sinogram, the evaluated region of interest was limited to the repaired area to avoid diluting the performance measurement. Additionally we applied a 3x3 Gaussian kernel [0.5 1 0.5] x [0.5 1 0.5] to both sinograms before comparison to filter out high frequency noise thereby providing a more realistic comparison. The normalized mean squared error (NMSE), which is an absolute measure, was calculated between the sinograms. That is,

\begin{align}
\text{NMSE}(\hat{y},y) &= \sum\limits_{i=0}^{k-1} (\hat{y}_i - y_i)^2 / \sum\limits_{i=0}^{k-1} y_i ^2
\label{sino_measure}
\end{align}

In the reconstructed image domain MS-SSIM, which uses a perception-based model, was used to compare the similarity between reference images and images reconstructed from sinograms with missing data and sinograms corrected with each repair method. In our experiment, all 876 images from the two test patient data sets were reconstructed from sinograms containing one, two, and four missing blocks in each slice and compared with images from ground truth sinograms. All reconstructions were performed using the standard Biograph mCT scanner's reconstruction software utilizing the Ordered Subset Expectation Maximization (OSEM) protocol with 3 iterations, 21 subsets and a 5x5 Gaussian filter. 
 
In the quantitative accuracy domain, three regions of interest were selected in the test data corresponding to the liver, uptake near the kidney and a small lesion. The mean tracer concentrations were measured across images based on ground truth, repaired and not repaired sinograms.

\section{Results}
  
\subsection{Sinogram Comparison}

\begin{figure*}[h]
 	\centering
 	\includegraphics[width=\columnwidth]{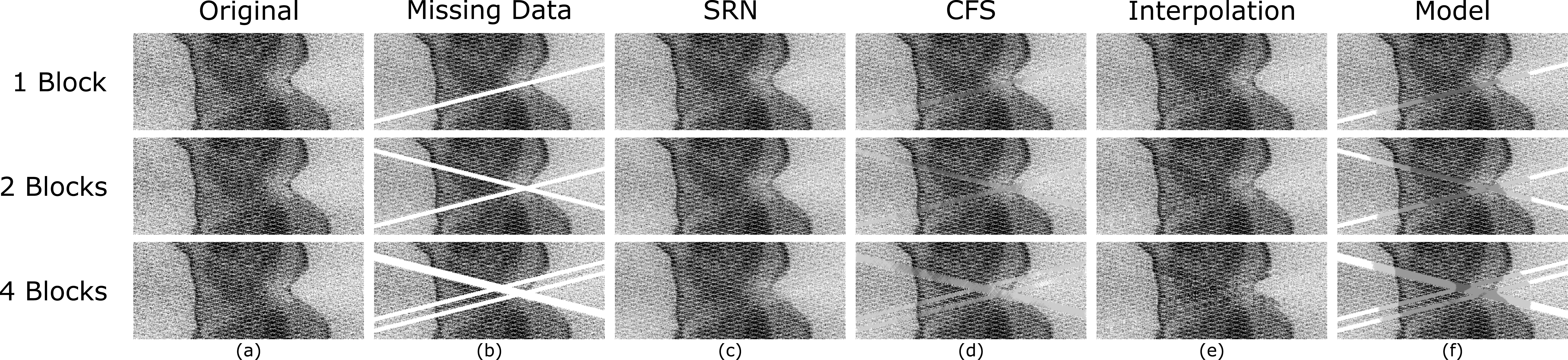}
 	\caption{A visual comparison of sinograms for each repair method across the range of missing data shows the sinograms corrected with the proposed neural network most closely resemble the ground truth compared to the other methods.
	}
 	\label{comparison}
\end{figure*}

Figure~\ref{comparison} provides a sample of repaired sinogram slices for each method along with the ground truth and the missing data. A visual comparison across all methods of repairing missing sinogram data show the established methods performing in accordance with previously published results, but the Sinogram Repair Network results most closely resemble the ground truth.

\begin{figure}[h]
	\centering
	\includegraphics[width=\columnwidth]{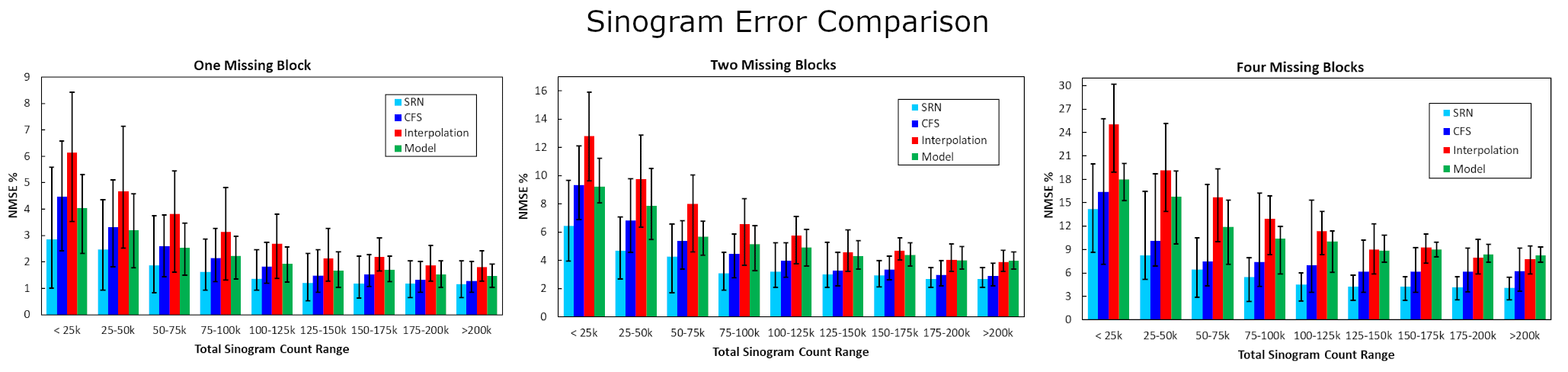}
	\caption{Plots of normalized mean squared error averaged over ten realizations of missing blocks with the maximum and minimum error shown for each method. The plots show that all repair methods perform the worst at low count densities and improve as density increases with the Sinogram Repair Network outperforming other methods across the spectrum of count density and number of missing block detectors.}
	\label{sinogram_slice}
\end{figure}

Figure~\ref{sinogram_slice} shows the normalized mean squared error from equation \ref{sino_measure} grouped into sinogram ranges 25,000 counts wide measured for each repair method averaged across ten random realizations of missing test set data for one, two and four missing blocks. Additionally the plots include bars to indicate the single slice maximum and minimum error for each method. Each of the repair methods has the largest error at low count densities. On inspection of these sinograms, the reason for the lower performance is the low signal to noise ratio in these regions, which tend to originate from either end of the whole-body scan where sensitivity is lower and in the case of the upper legs, the tracer distribution is more sparse. With this in mind, extra care must be taken when relying on sinogram repair in regions of high noise. As the count density increases however, all repair methods perform better and experience decreasing error and a narrowing of the difference between maximum and minimum error with the Sinogram Repair Network improving faster than the other methods with a widening performance gap especially as the amount of missing data increases. 

\subsection{Reconstructed Image Comparison}

\begin{figure}[ht]
	\centering
	\includegraphics[width=\columnwidth]{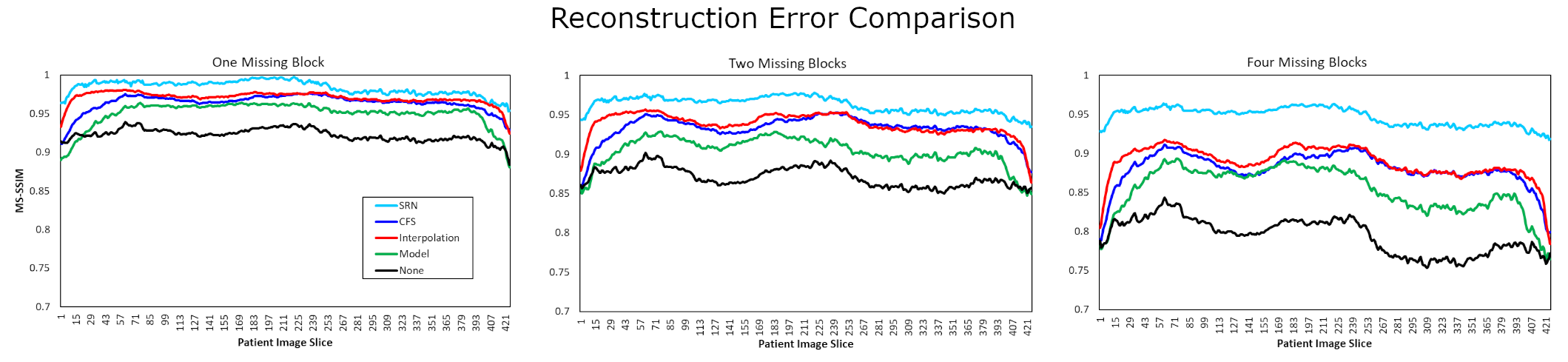}
	\caption{The multi-scale structural similarity measure of not repaired and repaired sinograms averaged across 10 random realizations of missing blocks compared to ground truth for one, two, and four missing blocks shows that in image space the previous repair methods perform approximately the same, while the Sinogram Repair Network achieves higher scores and is more robust to increased amounts of missing data.}
	\label{image_compare}
\end{figure}

Table \ref{ssim_results} and Fig.~\ref{image_compare} compare sinogram repair method performance in image space against ground truth using MS-SSIM for one, two and four missing blocks. The averaged performance across all test slices presented in Table \ref{ssim_results} shows that the previously published methods along with the Sinogram Repair Network all perform relatively well with only one missing block detector. However, as the amount of missing data increases, the performance gap between the previously established methods and the proposed neural network grew, which highlights the ability of this new method to handle higher quantities of failed block detectors. When the individual slices of a single patient from the test set are averaged over 10 realizations of missing block detectors, as shown in Fig.~\ref{image_compare}, the Sinogram repair network outperforms the other methods both across the range of whole-body images and quantities of missing data.   

Another observation of note is the lower performance of all repair methods on the start and end image planes, which we found to be consistent across data sets. These regions of a PET acquisition suffer from inherently low sensitivity and high noise due to a scanner's geometric constraints and thus produce both sparsely populated sinograms with fewer counts (oblique planes) and sinograms high noise components (direct planes). A visual inspection of repaired sinograms from this region confirms the relatively poor performance. Further research could be conducted to better understand the repair of sparse and noisy sinograms and develop more consistent techniques in these areas, but is beyond the scope of this study.

\begin{table}[h]
	\centering
	\caption{Results from the multi-scale structural similarity measure between reconstructed ground truth and repaired sinograms averaged over all sample images in the test data sets across 10 random realizations of missing detector blocks.}
	\begin{tabular}{||c|c|c|c|c|c||} 
		\hline
		& No Repair & Inter & CFS & Model & SRN \\ [0.5ex] 
		\hline\hline
		1 Missing Block  & 0.923 & 0.970 & 0.964 & 0.950 & \textbf{0.984}  \\ 
		\hline
		2 Missing Blocks & 0.871 & 0.940 & 0.932 & 0.914 &\textbf{0.963}  \\
		\hline
		4 Missing Blocks & 0.794 & 0.899 & 0.902 & 0.853 & \textbf{0.947}  \\
		\hline
	\end{tabular}
	
	\label{ssim_results}
\end{table}

\begin{figure*}[t]
	\centering
	\includegraphics[width=\columnwidth]{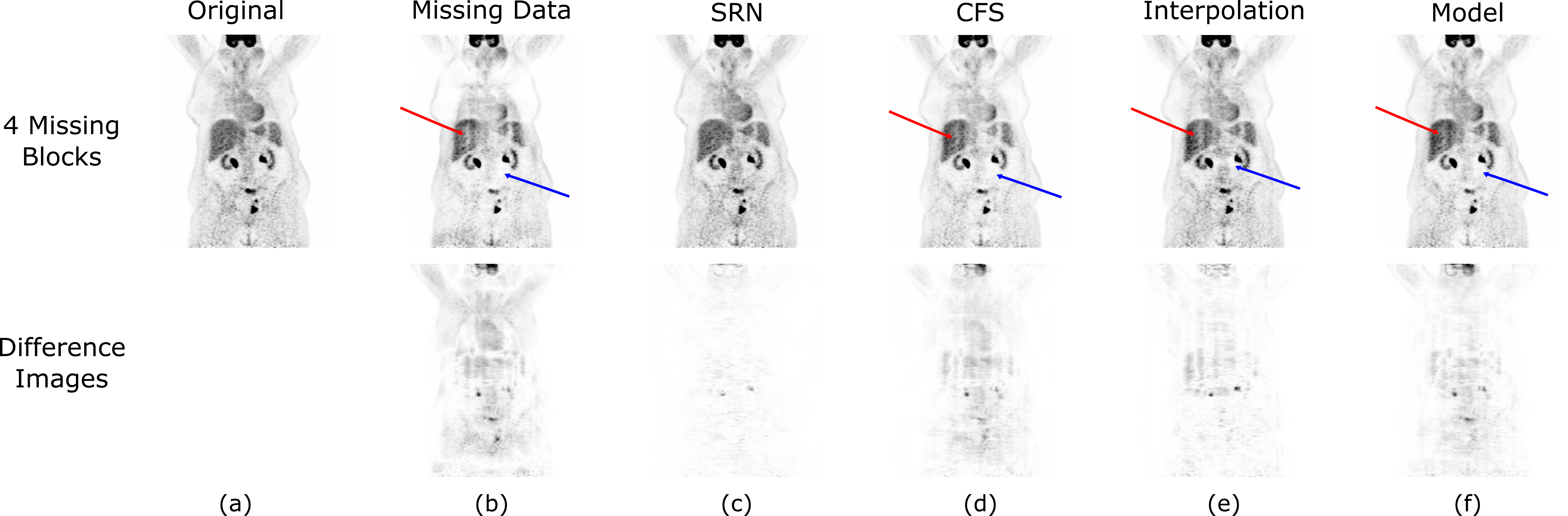}
	\caption{A reconstructed image space comparison of coronal slices from ground truth images and each of the repair methods with four simulated missing blocks. A visual inspection of the reconstructed and difference images show fewer artifacts with the Sinogram Repair Network. In comparison the other repair methods show evidence of streaks (red arrows) and low sensitivity (blue arrows).  
	}
	\label{imagespace_coronal}
\end{figure*}

Coronal slices of reconstructed image volumes with four missing blocks from the test data set are presented in Figure~\ref{imagespace_coronal} along with their difference image from ground truth. Visual inspection shows the Sinogram Repair Network reconstructed image is most similar to ground truth and has the least pronounced difference image. The images based on previous repair methods all show visual evidence of streaking as noted by the red arrow. They also contain areas of artificially low activity next to high activity as noted by the blue arrows. Both of these artifacts are confirmed by inspecting the difference image, but in contrast the Sinogram Repair Network image lacks these undesirable artifacts. The results from one and two missing blocks are not shown since differences in a single missing block are hard to distinguish visually and while images with two missing blocks do contain similar artifacts to the images shown, the visual differences are less pronounced.

\subsection{Quantitative Measurements} 
The use of quantitative measurements of tracer concentration in PET such as SUV \citep{kinahan:10} taken directly from the reconstructed image are used in treatment planning, clinical research, drug trials, and assessing response to therapy over multiple studies. It is therefore important when repairing sinograms to accurately represent the tracer concentration to avoid adverse affects on the analysis of patient images. 

\begin{figure}[t]
	\centering
	\includegraphics[width=\columnwidth]{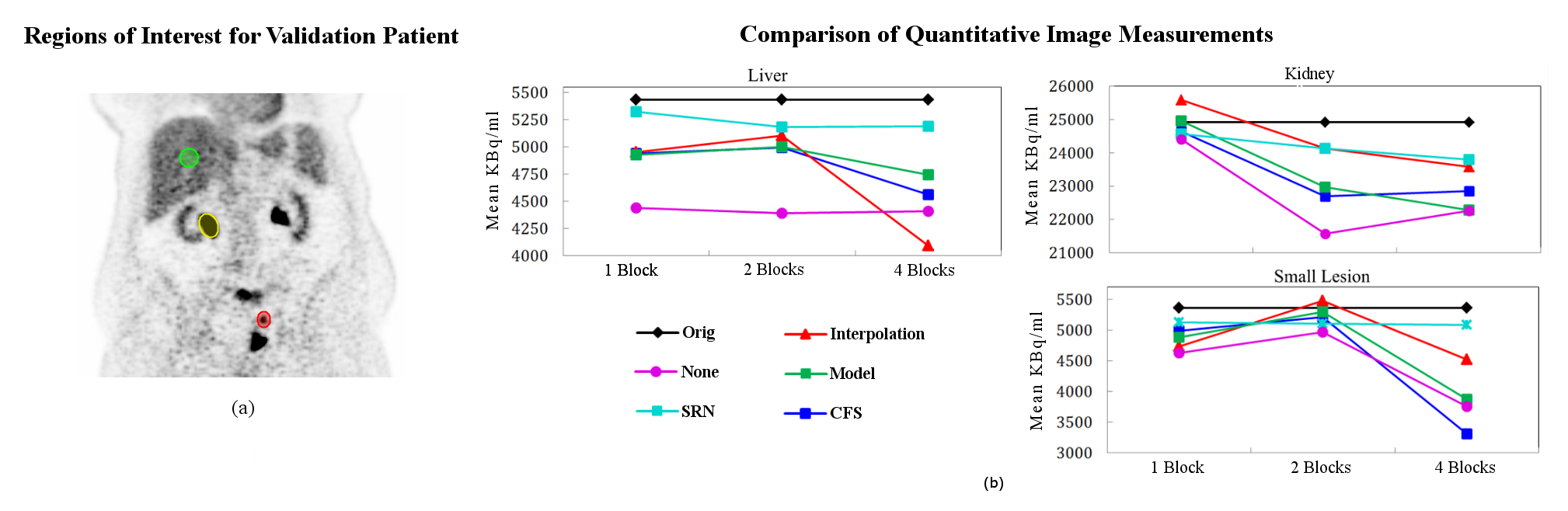}
	\caption{ROI comparision. (a) ROIs extracted from test data used to explore the impact of sinogram repair on quantitative PET measurements; yellow is in the area of the right kidney, red is a small lesion, and green is a liver measurement, which is often used as a reference. 
	(b) Quantitative results per ROI.
	}
	\label{quantitative}
\end{figure}

Figure \ref{quantitative}(a) shows the regions of interest (ROIs) examined in this study. The yellow ROI of the kidney measures 5,495~mm$^3$. The red ROI of a small lesion measures 929~mm$^3$ while the green ROI measures 2,892~mm$^3$ and is of the liver, which is commonly used as a reference point in quantitative PET measurements \citep{kinahan:10}. Fig.~\ref{quantitative}(b) gives a quantitative comparison of the repair methods across ROIs for 1, 2, and 4 missing blocks. Measurements from the original image reconstructed from the ground truth sinograms is shown in black and is considered the ideal for this comparison. As expected, the repair methods generally exhibit lower tracer concentration measurements than the original and this gap tends to widen as the number of simulated failed blocks increases. The Sinogram Repair Network maintains more consistent measurements across varying amounts of missing data and generally provides measurements closer to the ideal than other methods. Although the Sinogram Repair Network performs well, this study only examined quantitative measurements from two patients due to the limited data set and further work is necessary to draw strong and confident conclusions about quantitative reliability.

\section{Discussion}
\label{sec:discussion}

One application of the Sinogram Repair Network could be to continue clinical operations with full diagnostic quality in the event one or more block detectors fail the daily quality control procedure. This mitigation may become even more important as the number of block detectors and axial field of view of PET scanners continues to increase. However, not all detector failures have equal impact on the diagnostic quality of patient images; the number of bad block detectors and their relative location make a difference. The clinical relevance of bad blocks was first studied in 1999 \citep{Buchert:99}, and many of their suggestions are still prudent even with newer and more effective repair techniques. A key suggestion is to validate the diagnostic quality of the scanner by creating a mask of the failed detectors and then executing an extended quality control procedure by simulating the missing data on previously stored reference data sets representative of the clinical environment. These data sets are then repaired, reconstructed and scored using a structural similarity metric. The results of this metric along with a visual inspection and quantitative measurement comparison to the ground truth should determine if the diagnostic quality of the scanner is compromised and whether imaging should confidently continue or cease until a repair is made. For the most robust results the reference data sets should contain at least some small lesions at various distances from the center of the transaxial field of view.

\section{Conclusion}

The proposed Sinogram Repair Network is an effective method of repairing missing data in PET whole-body sinograms with one or more malfunctioning block detectors. A wide range of aspects were examined including sinogram and image space analysis and quantitative measurement accuracy and when compared with the existing repair methods, the Sinogram Repair Network was found to perform better in all cases.

Even with the superior performance of the sinogram repair method proposed in this paper, there are a number of avenues of further research that could provide improvements over the current performance level. One direction is to split the data set into smaller distributions based on some classifying feature such as the total counts in a sinogram. Then multiple networks could each be trained on a subset of the broader distribution and would be required to generalize to a much smaller portion of the global distribution of sinograms. Another promising direction is to consider 3D sinogram volumes when filling in missing data with the expectation that a neural network could learn the relationships between the various sinogram segments and utilize information from segments without missing data to fill gaps in affected sinograms. An expansion of this idea could also include the time-of-flight components present in most modern PET scanners.  Other avenues for further exploration include using alternative neural network structures such as the Wavelet Residual Network \citep{Kang:18} or the utilization of other loss functions such as those found in Generative Adversarial Networks \citep{Goodfellow:14}. 

Additionally, there are a number of other potential applications where the techniques in this research could be applied. One natural application for the proposed method is to fix normalization errors. We briefly experimented with failed detector blocks during continuous bed motion acquisitions, which causes artifacts similar to normalization errors, and found the sinogram repair method to work well. Perhaps the most impactful extension to our work is designing a PET/CT scanner with intentional gaps in the detector array (both transaxial and axial) and using techniques similar to those presented here to fill in the missing data creating a scanner with performance characteristics similar to a system without gaps in the detector array, but at a significantly lower cost, longer axial field of view and higher acquired data efficiency. This approach would enable all downstream image formation components to remain unchanged in contrast to other intentional gap techniques that require special reconstruction methods to effectively mitigate the gaps. 

\begin{small}
\newcommand{\newblock}{}
\bibliographystyle{apalike}
\bibliography{badblock}
\end{small}

\end{document}